\documentstyle[11pt]{article}

\title{\bf On the odderon intercept in the variational approach}
\author{M.A.Braun\\ Department of High Energy Physics,
University of St. Petersburg,\\ 198904 St. Petersburg, Russia.}
\date{January 1998}
\def\beq{\begin{equation}}
\def\eeq{\end{equation}}
\def\noi{\noindent}

\begin{document}
\maketitle
\medskip
\noi{\bf Abstract.}

 The odderon intercept
$ \alpha $ is recalculated by the
  variational method based
 on conformal invariance.  The
 final result is $\alpha=1-0.339\,\alpha_s N/\pi$ 
 in contrast to the published  value above unity.
\vspace{6.5cm}

\noi{\Large\bf SPbU-IP- 1998/3}
\newpage
\section{Introduction}
As shown in [1,2], at least in the high-colour limit, all states
formed by any number of reggeized gluons reduce to two basic ones: the
pomeron for the positive signature and the odderon for the negative
signature. The intercepts of these two states are fundamental for the
high-energy behavior of cross-sections in the perturbative QCD. Unlike
the pomeron intercept, known since more than two decades ago,
the odderon intercept has not been calculated exactly until now, despite
much activity around this problem and most sophisticated mathematical
tools used [3-6]. The only published results refer to estimates made in
the variational approach [7-9]. In paper [8], based on the conformal
property of the odderon [7], the obtained intercept lies above unity.
In the following we present it in the form
\beq
\alpha=1-\frac{\alpha_sN}{\pi}\frac{3}{2}\epsilon
\eeq
Here $\alpha_s$ is the strong coupling constant, $N$ is the number
of colours; $\epsilon$ has a meaning of the odderon "energy" per
gluon in units $\alpha_sN/\pi$. The corresponding energy for the pomeron
is $-4\ln 2$. It is negative, which implies the intercept above unity
and cross-sections growing with energy like $s^{\Delta}$ where
$\Delta=-(\alpha_sN/\pi)\epsilon$. For the odderon in [8] it was
obtained
\beq
\epsilon=-0.25
\eeq
which is also negative but considerably greater than the pomeron energy.
If correct, it would correspond to the negative signature cross-sections
also rising with energy, although much slowlier than the positive
signature ones.

In the subsequent paper [9] the conformal properties of the odderon were
not taken into account. Instead its wave function was presented as a
sum of one-gluon "orbitals", which made it possible to take a very
large number of variational parameters. A positive value was obtained
for the energy:
\beq
\epsilon=0.298
\eeq
In principle, due to the variational nature of the calculations, this
value does not contradict (2). However in [9] certain theoretical
arguments were also presented in favour of the odderon intercept being
smaller than unity.

Guided by the evident conflict of these theoretical arguments with the result
(2), we undertook to repeat the variatuional calculations in the
conformal approach of [7,8], taking both the same trial function and the
ones with more parameters. It turned out that the result (2) is in error.
Correct
calculations with the same trial function (with 2 parameters) give \beq
\epsilon=0.2286\pm 0.0001
\eeq
with an opposite sign as compared to (2). Inclusion of more parameters
(up to 8) allows to slightly lower this value down to
\beq
\epsilon=0.2262\pm 0.0001
\eeq
The uncertainties in (4) and (5) refer to errors in numerical
calculations. In view of a very small change in $\epsilon$ upon
introducing more parameters, one is inclined to take (5) as a final value
for the odderon energy with a precison of 0.1\%.

The study of the odderon energy in the conformal approach, in fact,
presents a formidable calculational problem, deserving some attention.
In this
note we briefly discuss the difficulties involved and some intermediate
steps in obtaining our results.

\section{Basic equations}
 Basic equations for the odderon energy in the conformal
approach were derived in [7]. We reproduce them here to fix our
notatons. The odderon energy is to be sought as a minimum of the energy
functional
\beq
E=\sum_{n=-\infty}^{\infty}\int_{-\infty}^{\infty}d\nu\epsilon_{n}(\nu)
|\alpha_{n}(\nu)|^{2}
\eeq
Here
\[
\alpha_{n}(\nu)=\int_{0}^{\infty}dr r^{-2-2i\nu}\int_{0}^{2\pi}
d\phi e^{-in\phi}\]\beq
\left(i\nu+\frac{n+1}{2}+re^{i\phi}(h-i\nu-\frac{n-1}{2})\right)
(i\nu-\frac{n-1}{2})(-\tilde{h}+i\nu-\frac{n-1}{2})Z(r,\phi)
\eeq
$Z$ is the odderon wave function
and
\beq
\epsilon_{n}(\nu)=2\ {\mbox Re}\,\left(\psi(\frac{1+|n|}{2}+i\nu)
-\psi(1)\right)
\eeq
The normalization condition is
\beq
1=D=\sum_{n=-\infty}^{\infty}\int_{-\infty}^{\infty}d\nu
|\alpha_{n}(\nu)|^{2}
\eeq

In these formulas
$ h $ and
$ \tilde{h} $ are the two conformal weights, which are taken
to be equal to 1/2. The trial function
$ Z $ then takes the form
\beq
Z(r,\phi)=r^{1/3}r_{1}^{1/3}z(r,\phi)
\eeq
where \beq r_{1}^{2}=1+r^{2}-2r\cos\phi
\eeq
and  $z(r,\phi)=z(r,r^*)$ shoud be invariant under the transformations
of complex $r$: $r\rightarrow -1/r$ and $r\rightarrow 1-r$. This can be
achieved by taking $z=z(a)$ where
\beq
a=\frac{r^{2}r_{1}^{2}}{(1+r^{2})(1+r_{1}^{2})(r^{2}+r_{1}^{2})}
\eeq

We take the trial function $z(a)$ in the form
\beq
z(a)=\sum_{k=1}^{N-1}c_ka^{k/2-1/6}+c_Na^{5/6}\ln a
\eeq
with $N-1$ variational parameters $c_k$ ( one of $c_k$ is determined
by the normalization condition). For $N=3$ this function coincides
with the one used in [8].

The basic quantity
$ \alpha $ given by Eq. (7) can  be presented in the
 form
\beq
\alpha_{n}(\nu)=(\frac{n}{2}-i\nu)
(\frac{n-1}{2}-i\nu)\big[(\frac{n+1}{2}+i\nu)f_{n}^{(1)}(\nu)+
 (1-\frac{n}{2}-i\nu)
f_{n-1}^{(0)}(\nu)\big]
\eeq
where
\beq
f_{n}^{(k)}=\int_{0}^{\infty}dr r^{-1-k-2i\nu}\int_{0}^{2\pi}d\phi
e^{-in\phi}Z(r,\phi)
\eeq

Function
$ f_{n}^{(k)}(\nu) $ has  the following properties.
Since
$ Z $ is real,
$ (f_{n}^{(k)}(\nu))^{*}= f_{-n}^{(k)}(-\nu)$
and since
$ Z $ only depends on
$ \phi $ through
$ \cos\phi $, we have
 $f_{n}^{(k)}(\nu)= f_{-n}^{(k)}(\nu)$.
 Under
$ r\rightarrow 1/r $ the argument
$ a$  stays invariant, from which it follows that
$ Z\rightarrow Z/r $. As a result we obtain
$ f_{n}^{(1)}(\nu)= f_{n}^{(0)}(-\nu)=(f_{n}^{(0)})^{*}$.
Using these  properties we can restrict  the summation over $n$ and
 integration over
$ \nu $ to nonnegative values. 
 The appearing coefficient 2 
 cancels in dividing $ E $ by
$ D $  so that we need not take it into account. The
value of the
$|\alpha_{n}(\nu)|^{2}
 $ can evidently be expressed via a single
function
$ f_{n}^{(0)}(\nu) $, which will simply be denoted  as
$ f_{n}(\nu) $ in the following. In this manner we obtain our basic
formulas for $ E $ and $ D $:
\beq
E=\sum_{n=0}^{\infty}\int_{0}^{\infty}d\nu\epsilon_{n}(\nu)
p_{n}(\nu)
\eeq
where for
$ n>0 $
\[
p_{n}(\nu)=(\frac{n^{2}}{4}+\nu^2)(\frac{(n-1)^{2}}{4}+\nu^2)\big[
(\frac{(n+1)^{2}}{4}+\nu^2)|f_{n}(\nu)|^{2}+
(\frac{(n-2)^{2}}{4}+\nu^2)|f_{n-1}(\nu)|^{2}+\]\beq
2\,{\mbox
Re}\,(\frac{n+1}{2}-i\nu)(\frac{2-n}{2}-i\nu)f_{n}(\nu)f_{n-1}(\nu)\big]
\eeq
and
\beq
p_{0}(\nu)=\nu^2(\frac{1}{4}+\nu^2)[(\frac{1}{4}+\nu^2)|f_{0}(\nu)|^{2}+
(1+\nu^2)|f_{1}(\nu)|^{2}+2\,{\mbox
Re}\,(\frac{1}{2}-i\nu)(1-i\nu)f_{0}(\nu)f_{1}(\nu)]
\eeq
The normalization functional
$ D $ has the same form (16) with
$ \epsilon_{n}(\nu)\rightarrow 1 $. Thus calculation of the
odderon energy requires calculation of functions
$ f_{n}(\nu) $ and
$ \epsilon_{n}(\nu) $.

The main technical difficulty 
is a double Fourier transform (15). The energy
$\epsilon_n(\nu)$ in $E$, Eq. (16), monotonously grows both with
$n$ and $\nu$. It is negative only for $n=0$ and small enough values of
$\nu$. So the problem with this formalism is that cutting in (16)
summation over $n$ and integration over $\nu$ by some maximal values
$n_m$ and $\nu_m$, one always gets smaller $E$ than the exact value,
corresponding to $n_m$ and $\nu_m\rightarrow\infty$. Therfore in the
course of the calculation one always approaches to the variational value
of $\epsilon$ {\it from below}.
 As we shall see in the following, in fact,
rather high values of $n_m$ and $\nu_m$ are necessary to obtain
$\epsilon$ with some degree of accuracy
\footnote{This is the reason why in [7]
the negative value (2) was obtained. The authors chose too small
values of $n_m$ and $\nu_m$}. On the other hand, with high $n$ and
$\nu$,
the double Fourier transform (15) becomes very difficult, especially
having in mind that, due to the factors in (17), two first terms in the
asymptotic expansion of $\alpha$ at high $n$ and $\nu$ cancel.
As a result, as mentioned in the Introduction, a trustworthy calculation
of $E$ and $D$ turns out to be very complicated, in spite of it
superficial transparency.

The crucial point in obtaining reasonable results has been using
analytic
asymptotic expansions for $f$ at high $n$ and $\nu$, which are discussed
in the next section.

\section{Asymptotics at large $n$ and $\nu$}
Passing to variable
$\rho=-\ln r$ and introducing 2-dimensional vectors $x=(\rho,\phi)$.
and $w=(z,n)=(2\nu,n)$ we rewrite (15) as
\beq f_n(\nu)=\int_{-\infty}^{+\infty}d\rho\int_{-\pi}^{+\pi}d\phi
e^{iwx}Z(x)\eeq
Evidently for the asymptotics at high $n$ and $\nu$ the integration
point $x=0$ is essential.

At $x\rightarrow 0$, keeping terms up to third order in small $\rho$ and
$\phi$, we have
\[r^{1/3}=1-(1/3)\rho+(1/18)\rho^2-(1/162)\rho^3\]
\[r_1^{1/3}=x^{1/3}(1-(1/6)\rho+(1/36)\rho^2-(1/72)\phi^2-(1/324)\rho^3+
(1/432)\rho\phi^2)\]
\[a^p=(1/2)^p x^{2p}(1-p(29/12)\rho^2-p(25/12)\phi^2)\]
(no terms of the third order appear in $a$). In these formulas
$x=\sqrt{\rho^2+\phi^2}$. As a result,  at small $x$
\beq Z_p=(rr_1)^{1/3)}a^p=(1/2)^p x^{2p+1/3}(1-(1/2)\rho
+a\rho^2+b\phi^2+c\rho^3+d\rho\phi^2)\eeq where
\[a=5/36-(29/12)p;\ \ b=-1/72-(25/12)p;\ \ c=-1/36+(29/24)p;\ \
d=1/144+(25/24)p\]
For the term with a logarithm, in the same manner we obtain
\[\tilde{Z}_p=(rr_1)^{(1/3)}a^p\ln a=(1/2)^p x^{2p+1/3}[\ln
(x^2)(1-(1/2)\rho +a\rho^2+b\phi^2+c\rho^3+d\rho\phi^2)+\]\beq
\ln(1/2)-(1/2)\ln(1/2)\,\rho+a_1\rho^2+b_1\phi^2+c_1\rho^3+
d_1\rho\phi^2)]\eeq
where
\[f_1=f\ln(1/2)+\delta_f,\ \ f=a,b,c,d\]
and
\[\delta_a=-29/12,\ \delta_b=-25/12,\ \delta_c=29/24,\ \delta_d=25/24\]

Inserting these expressions into the integral (19) and extending the
integration over $\phi$ to the whole real axis one obtains the
asymptotical expansion of different terms in $f_n(\nu)$. In
particular the asymptotical expansion of the term originating from
$Z_p$ is found as
\[f_n^{(p)}(\nu)=c_p(1+(1/2)id/dz-ad^2/dz^2-bd^2/dn^2+icd^3/dz^3+id
d^3/dzdn^2)w^{-\alpha-1}\]
where $\alpha=2p+4/3$ and
\[c_p=2^{\alpha+1-p}\Gamma^2(1/2+\alpha/2)\cos (\pi\alpha/2)\]
Doing
the derivatives, one obtains finally
\beq {\rm Re}
f_n^{(p)}(\nu)=c_pw^{-\alpha-1}\left(1+(\alpha+1)w^{-4}[z^2(b-a(\alpha+2)
+n^2(a-b(\alpha+2))]
\right)\eeq
\beq {\rm Im}
f_n^{(p)}(\nu)=-(1/2)c_p(\alpha+1)zw^{-\alpha-3}\left(1
+2(\alpha+3)w^{-4}[z^2(c(\alpha+2)-d)+n^2(d(\alpha+4)-3c)]\right)\eeq

For the term with a logarithm at $p=5/6$ only the part with $\ln x^2$
contributes. The result coincides with the formula above with $p=5/6$ and the
constant $c_p$ substituted by
$16\pi (1/2)^{5/6}$.

If one puts these asymptotic expressions into (17) one finds that the two
leading terms coming from $Z_{1/3}$ cancel. Numerically the asymptotic
expansion begins to work at rather high values of $n$ and $\nu$:
$\sqrt{n^2+4\nu^2}>30$.

\section{Numerical procedure}
According to (15) function
$ f_{n}(\nu) $  is defined as an integral
\beq
f_{n}=\int_{0}^{\infty}dr r^{-1-2i\nu}\int_{0}^{2\pi}d\phi
e^{-in\phi}Z(r,\phi)
\eeq
The double Fourier transform  (24) was done numerically in the interval
$0\leq n<30$, $0<\nu<15$.

At $r>1$ we transform the integration variable
$ r\rightarrow 1/r $, under which
$ Z\rightarrow Z/r $.  
Then the integration over
$ r $  becomes restricted to the interval $ 0<r<1 $.
Using properties of
$ Z $ we can also integrate only over
$ 0<\phi<\pi $ and substitute $ e^{-in\phi}
$  by
$ 2\cos n\phi $. We finally use that
$ Z $ linearly goes to zero as
$ r\rightarrow 0 $ and introduce a new function
\beq
\zeta(r,\phi)=Z(r,\phi)/r
\eeq
In terms of
$ \zeta $  we then have
\beq
f_{n}=2\int_{0}^{1}dr
(r^{-2i\nu}+r^{-1+2i\nu})\int_{0}^{\pi}d\phi
\cos in\phi\ \zeta(r,\phi) \eeq

The integrand is evidently singular at
$ r=0 $.  To soften this
 behaviour
 we use that at small $r$
\beq
\zeta(r,\phi)=\zeta_{0}-\zeta_{1}r\ln r+\zeta_{2}r+{\cal
O}(r^{2}\ln r)
\eeq
where
\beq
\zeta_{0}=c_1(1/2)^{1/3},\ \zeta_{1}=-2c_{N}(1/2)^{5/6},\
\zeta_{2}=(1/2)^{5/6}(c_{2}-c_N\ln 2)
\eeq
In (26) we subtract from 
$ \zeta $ the first term of expansion (27) in the part of the
integrand with
$ r^{-2i\nu} $ and the first three terms of this expansion
in the part with
$ r^{-1+2i\nu} $, adding to $ f_{n}(\nu) $ the result of the
integration of subtracted terms, which can be found
explicitly. In this way we come to our final formula \[
f_{n}(\nu)=2\int_{0}^{\pi}d\phi
\cos in\phi\int_{0}^{1}dr
\big(r^{-2i\nu} (\zeta(r,\phi)-\zeta_{0})\]\[
+r^{-1+2i\nu}(\zeta(r,\phi)-\zeta_{0}+\zeta_{1}r\ln
r-\zeta_{2}r)\big)+\]\beq\delta_{n0}2\pi
(\frac{\zeta_{0}}{2i\nu(1-2i\nu)}+
\frac{\zeta_{1}}{(1+2i\nu)^{2}}+\frac{\zeta_{2}}{1+2i\nu}
\eeq

Eq. (29) was used for numerical calculation of
$ f_{n}(\nu) $ in the above mentioned interval of $n$ and $\nu$.
Integrations were performed by dividing the rectangle
$0<r<1,\ 0<\phi<\pi$ into an $M\times M$ grid, interpolating $\zeta$
quadratically on the grid and then doing the integrals explicitly.
The maximal value of $M$ was 640. The achieved accuracy was about $10^{-5}$.

Thus calculated values of $f_n(\nu)$ were summed over $n$ and integrated
over $\nu$ as indicated in (16)-(18) to obtain $E$ and $D$.
Stable results were
obtained with the quite high maximal values $n_m=300$ and $\nu_m=150.$.
In the part of $n,\nu$ space outside the rectangle
$0\leq n<30,\ 0<\nu<15.$
 the asymptotic expressions (22) and (23) were used for $f_n(\nu)$.

As a result we calculated $E$ and $D$ as a quadratic form in the variational
parameters $c_k$. Afterwards the minimal value $\epsilon$ of $E$ subject
to condition $D=1$ was found by standard methods.

The results for different number of parameters $N$ are presented in the Table.
To see the importance of the high $n,\nu$ region, we also present the values
of energy $\epsilon_1$ calculated without the asymptotical region, that is
restricting to $0\leq n<30,\ 0<\nu<15.$. The standard precision corresponds
to the $r,\phi$ grid 320$\times$320. To clarify the accuracy achieved we also
present the results with a double precision (the grid 640$\times$640) for
$N=5$.

Inspecting these results we see that the final accuracy in energy
is of the order $7.10^{-5}$.
This implies that taking $N>6$ has no sense within the
precision achieved, since the corresponding change in energy is  of the
same order or less.

So our conclusion is that the variational odderon energy with a trial
function (13) is given by (5) and  that with the accuracy  achieved in the
course of numerical integration, as described above, the maximal number
of terms to be taken in the trial function is $N=6$, although already with
$N=3$ used in [8] the energy is obtained up to 1\%.

\section{Acknowledgements}
The author expresses his deep gratitude to P.Gauron, L.N.Lipatov and
B.Nicolescu for stimulating discussions and N.Armesto for fruitful
collaboration.

\section{References}
1. M.A.Braun, Z.Phys. {\bf C71} (1996) 123.\\
2. M.A.Braun and G.P.Vacca, Bologna univ. preprint, hep-ph/9711486.\\
3. L. N. Lipatov, JETP Lett. {\bf 59} (1994) 571.\\
4. L. D. Faddeev and G. P. Korchemsky, Phys. Lett. {\bf B342} (1995) 311.\\
5. G. P. Korchemsky, Nucl. Phys. {\bf B443} (1995) 255;
{\bf B462} (1996) 333.\\
6. J.Wosiek and R.A.Janik, preprint TPJU-21/96 (hep-th/9610208).\\
7. P.Gauron, L.Lipatov and B.Nicolescu
Phys.Lett. {\bf B260} (1991) 407.\\
8. P.Gauron, L.Lipatov and B.Nicolescu, Phys. Lett. {\bf B304} (1993) 334;
Z.Phys. {\bf C63}(1994)253.\\
9. N.Armesto and M.A.Braun, Z.Phys. {\bf C75} (1997) 709.\
\newpage
\section{Table}
\begin{center}
{\bf Odderon energy and parameters of the trial function}
\vspace{1.cm}

\begin{tabular}{|r|c|l|c|}
\hline
N&$\epsilon$&$c_2,c_3,...(c_1=1)$&$\epsilon_1$\\\hline
3&0.22865&-0.5036,0.2895 &0.21990\\\hline
4&0.22632&-0.2791,-0.3190,0.3609&0.20663\\\hline
5&0.22627&-0.2009,-0.5052,0.1557,&0.17919\\
 &       & 0.3779                &        \\\hline
$5^*$&0.22634&-0.2021,-0.5028,0.1543,&       \\
  &        &0.3775                 &       \\\hline
6& 0.22619&-0.3735,0.08842,-0.9765,&-0.0015803\\
 &        &1.003,0.3471            &        \\\hline
7& 0.22618&                        &-0.37771\\\hline
8& 0.22616&                        &-0.54117\\\hline
9& 0.22616&                        &-0.60273\\\hline
\end{tabular}
\end{center}
\hspace{3.cm}$^*)$ double precision.
\end{document}